\documentclass[aps,showpacs, nofootinbib]{revtex4}
\newcommand{\be}{\begin{equation}}
\newcommand{\ee}{\end{equation}}
\newcommand{\ben}{\begin{eqnarray}}
\newcommand{\een}{\end{eqnarray}}

\newcommand{\cO}{{\cal O}}

\newcommand{\p}{\partial}
\newcommand{\na}{\nabla}

\newcommand{\tpsi}{\tilde \psi}
\newcommand{\tim}{{\tilde \mu}}
\newcommand{\tom}{{\tilde \omega}}

\newcommand{\tM}{{\tilde M}}

\newcommand{\ep}{\epsilon}

\newcommand{\ga}{\gamma}


\begin{document}

\title{Decay of Massive Scalar Hair on Brane Black Holes}
%%%%%%%%%%%%%%%%%%%%%%%%%%%%%%%%%%%%%%%%%%%%%%%%%%%%%%%%%%%%%%
%\author{David Langlois}
%\affiliation{Institute of  \protect \\
%University \protect \\
%20-031 , France \protect \\
%david@tytan.umcs.lublin.fr \protect \\

\author{Marek Rogatko and Agnieszka Szyp\l owska}
\affiliation{Institute of Physics \protect \\
Maria Curie-Sklodowska University \protect \\
20-031 Lublin, pl.~Marii Curie-Sklodowskiej 1, Poland \protect \\
rogat@tytan.umcs.lublin.pl \protect \\
rogat@kft.umcs.lublin.pl}

%%%%%%%%%%%%%%%%%%%%%%%%%%%%%%%%%%%%%%%%%%%%%%%%%%%%%%%%%%%%%%%%%%%%
\date{\today}
%\pacs{04.30.Nk, 04.40.-b}

%%%%%%%%%%%%%%%%%%%%%%%%%%%%%%%%%%%%%%%%%%%%%%%%%%%%%%%%%%%%%%%%%%%%%%%%%%%%%%%%%%%%%%%%%%%%%%%%%%%
\begin{abstract}
We study analytically the intermediate and late-time behaviour of the massive 
scalar field in the background of static spherically symmetric brane black hole solutions.
The intermediate asymptotic behaviour of scalar 
field reveals the dependence on the field's parameter mass as well as the multiple moment $l$,
while the late-time behaviour has the power law decay rate independent on those factors.
\end{abstract}
%%%%%%%%%%%%%%%%%%%%%%%%%%%%%%%%%%%%%%%%%%%%%%%%%%%%%%%%%%%%%%%%%%%%%%%%%%%%%%%%%%%%%%%%%%%%%%%%%%%%

\maketitle

%%%%%%%%%%%%%%%%%%%%%%%%%%%%%%%%%%%%%%%%%%%%%%%%%%%%%%%%%%%%%%%%%%%%%%%%%%%%%%%%%%%%%%%%%%%%%%%%%%%%%
\section{Introduction}
Recently there has been growing interests in studying models which basic idea is that our universe 
is only a submanifold on which the standard model is confined to inside a higher dimensional
spacetime. In the above model only geometric degrees of freedom can propagate in extra dimensions. 
By making the volume of extra dimensions spacetime large, one was able to lower a fundamental quantum
gravity scale to the electrovac scale of TeV-order.
One can also construct the black hole solutions in the brane-world models. 
The difficulties in those attempts stem from the fact that in general brane dynamics 
generate Weyl curvatures which in turn backreact on the brane dynamics.
\par
In order to simplify this attitude one can look for the analytic solutions to the projected Einstein equations
on the brane Ref.\cite{dad00}. Similar kind of solutions were revealed also in Refs.\cite{cas02}.
On the other hand, it is interesting to pose a question whether a brane on which four-dimensional
black hole is situated can be found by looking for a slice that intersects a bulk black hole.
It was revealed \cite{kod02},\cite{tan03} that
brane solutions with a black hole geometry cannot be found by slicing a bulk with $G(D-2,k)$ symmetry
if the brane is vacuum and not totally geodesic. In Ref.\cite{sea05} a localized static but non-vacuum brane 
black hole solution of a slice of a $G(D-2,k)$ bulk was presented. Recently, in Ref.\cite{gal06}
the possibility of finding a regular  Randall-Sundrum (RS) brane world on which a static spherically symmetric
black hole surrounded by realistic matter is located by slicing a fixed five-dimensional bulk black hole 
spacetime was presented.
\par
The decay of black hole hair is a very interesting problem on its own.
Late-time behaviour of various fields in the spacetime of a collapsing
body is of a great importance for  black hole's physics due to the fact that 
regardless of details of the collapse or the structure and properties
of the collapsing body the resultant black hole can be described only by few parameters
such as mass, charge and angular momentum, {\it black holes have no hair}. 
Price in \cite{pri72} for the first time studied the neutral external perturbations and found 
that the late-time behavior is dominated by the factor $t^{-(2l + 3)}$, for each 
multipole moment $l$. For the decay along null infinity and along the
future event horizon it was found Ref.\cite{gun94} that
the power laws are of the forms $u^{-(l + 2)}$ and $v^{-(l + 3)}$, where $u$ and $v$ were
the outgoing Eddington-Finkelstein (ED) and ingoing ED coordinates.
In Ref.\cite{bic72} the scalar perturbations on Reissner-Nordtr\"om (RN)
background for the case when  $\mid Q \mid < M$ was studied and it was shown the following dependence on time $t^{-(2l + 2)}$,
while for $\mid Q \mid = M$ the late-time behavior at fixed $r$ is governed by
$t^{-(l + 2)}$. 
Charged hair
decayed slower than a neutral one \cite{pir1}-\cite{pir3}, while
the late-time tails in gravitational collapse of a self-interacting
fields in the background of Schwarzschild solution was reported by Burko \cite{bur97}
and in  
RN solution at intermediate late-time was considered
in Ref.\cite{hod98}. 
The very late-time tails of the massive scalar fields in the Schwarzschild and nearly extremal RN black holes
were elaborated in Refs.\cite{ja}, \cite{ja1}. It was revealed that 
the oscillatory
tail of scalar field has the decay rate of $t^{-5/6}$ at                                   
asymptotically late time.
The power-law tails in the evolution of a charged massless scalar field around a fixed
background of dilaton black hole was studied in Ref.\cite{mod01a}, while the case of massive
scalar field was treated in \cite{mod01b}.
The problem of the late-time behaviour of massive Dirac fields were studied
in the spacetime of Schwarzschild black hole \cite{jin04}, while in the spacetime of
RN black hole was analyzed in Ref.\cite{jin05}. 
\par
The intense growing of interests in unification scheme such as superstring/M-theory triggered also the interests in
hair decays in the spacetimes of $n$-dimensional black holes.
As far as the  $n$-dimensional static black holes is concerned, the
{\it no-hair} theorem for them
is quite well established \cite{unn}. The  mechanism of decaying black hole hair in higher dimensional static
black hole case concerning the evolution of massless scalar field in the $n$-dimensional Schwarzshild
spacetime was determined in
Ref.\cite{car03}. It was found that for odd dimensional spacetime the field  decay had a
power falloff like $t^{-(2l + n - 2)}$, where $n$ is the dimension of the spacetime.
This tail was independent of the presence of the black hole.
For even dimensions the late-time behaviour is also in the power law form but in this case it is due to
the presence of black hole $t^{-(2l + 3n - 8)}$.
%%%%%%%%%%
The late-time tails of massive scalar fields in the spacetime of $n$-dimensional 
static charged black hole was elaborated in Ref.\cite{mod05}, where because of tremendous difficulties 
in solving analytically differential Eqs. of motion for the field numerical simulations for the spacetime dimension
$n = 5$ and $n = 6$ were performed.
\par
The main purpose of our paper will be to clarify what kind of mass-induced behaviours play the dominant role
in the asymptotic late-time tails as a result of decaying the massive scalar hair in the background of brane black hole.
The intermediate and late-time tails of the fields corresponding to the massive scalar ones have not been 
studied before in the context of brane black hole physics. On the other hand, our purpose is to see what effects 
of brane black hole parameters will have on the decay of the massive scalar hair on them.
\par
The paper is organized as follows.
In Sec.II we gave the analytic arguments concerning
the decay of scalar massive hair in the background of the considered black holes. 
Sec.III will be devoted to a summary and discussion.

%%%%%%%%%%%%%%%%%%%%%%%%%%%%%%%%%%%%%%%%%%%%%%%%%%%%%%%%%%%%%%%%%%%%%%%%%%%%%%
\section{The Decay of Scalar Hair in the Background of Black Hole Brane Solution}
\subsection{Casadio-Fabbri-Mazzacurati (CFM) brane black hole solution}
In Ref.\cite{cas02} it was revealed that 
projecting the vacuum $n + 1$-dimensional Einstein equations on a timelike manifold of
codimension one leads us to the analogs of the momentum and {\it Hamiltonian} constraints in the 
Arnowitt-Deser-Misner (ADM) decomposition of the metric. Now their role is to select our admissible field
configurations along hypersurfaces of constant coordinate. Further, it was argued that
this {\it Hamiltonian} constraint is a weaker requirement than the purely $n$-dimensional
vacuum Eqs. and it is equivalent to the relation for $n$-dimensional Ricci tensor $R_{ij} = E_{ij}$ ,
where $E_{ij}$ is the projection of $n + 1$-dimensional Weyl tensor on the brane \cite{shi00}.
\par
Our general setting will be brane black hole solutions proposed in Ref.\cite{cas02}. Two families of 
analytic spherically symmetric solutions with the condition that $g_{rr} \ne - {1 \over g_{tt}}$, 
parametrized by ADM mass and the post-Newtonian (PPN) parameter $\beta$ were found.
The parameter $\beta$ affects the perihelion shift and the Nordtvedt effect \cite{wil93}. \\
In our considerations we shall use the general form of spherically symmetric line element
\be
ds^2 = - A(r) dt^2 + {1 \over B(r)}dr^2 + r^2 d\Omega^2.
\ee
The momentum constraints are identically satisfied by the metric coefficients and the {\it Hamiltonian}
constraints can be written out for them \cite{cas02}. Setting $A(r) = \bigg( 1 - {2 M \over r} \bigg)$ the resulting
metric yields
\be
ds^2 = - \bigg( 1 - {2 M \over r} \bigg) dt^2 +
{\bigg( 1 - {3 M \over 2 r} \bigg) \over
\bigg(1 - {2 M \over r} \bigg)~\bigg( 1 - {\gamma M \over 2 r} \bigg)} dr^2
+ r^2 d\Omega^2,
\ee
where $\gamma = 4 \beta -1$. One can see that for $\beta = 1$ this solution reduces to the Schwarzschild 
black hole solution.
%%%%%%%%%%%%%%%%%%%%%%%%%%%%%%%%%%%%%%%%%%%%%%%%%%%%%%%%%%%%%%%%%%%%%%%%%%%%%%%%%
Next to find the pattern of the decay of massive scalar hair in the underlying brane black hole spacetime
we define the tortoise coordinates $y$ as
\be
dy = {dr \over 
A(r)^{1 \over 2}~B(r)^{1 \over 2}}.
\ee
This change of variables enables one to rewrite the line element in the following form:
\be
ds^2 = A(r) \bigg( -dt^2 + dy^2 \bigg) + r^2 d\Omega^2.
\ee
As we shall consider the scalar massive wave equation
\be
\bigg( \na_{\mu} \na^{\mu} - m^2 \bigg)\tpsi = 0,
\ee
in the spherically symmetric background it will be not amiss to resolve the field into spherical harmonics
\be
\tpsi = \sum_{l,\delta} {1 \over r} \psi_{\delta}^{l}(t, r) 
Y_{l}^{\delta}(\theta, \phi),
\ee
where $Y_{l}^{\delta}$ is a scalar spherical harmonics on the unit two-sphere. It leads to
the following equations of motion for each multipole moment
\be
\psi_{,tt} - \psi_{,yy} + V \psi = 0,
\label{mo}
\ee
where the effective potential $V$. 
\par
We shall analyze the time evolution of massive scalar field in the background of
brane black hole by means of the spectral decomposition method.
As was claimed in Refs.\cite{hod98},\cite{lea86} it was shown that the asymptotic tail is connected with the
existence of a branch cut situated along the interval $-m \le \omega \le m$.
An oscillatory inverse power-law behaviour of massive scalar field arises
from the integral of Green function $\tilde G(y, y';\omega)$ around branch cut.
In our paper we denote it by $G_{c}(y,y';t)$ and our main aim will be to find analytical form
of its integral.\\
The time evolution of massive scalar field may be written in the following form:
\be
\psi(y, t) = \int dy' \bigg[ G(y, y';t) \psi_{t}(y', 0) +
G_{t}(y, y';t) \psi(y', 0) \bigg],
\ee
for $t > 0$, where   the Green's function  $ G(y, y';t)$ is given by the relation
\be
\bigg[ {\p^2 \over \p t^2} - {\p^2 \over \p y^2 } + V \bigg]
G(y, y';t)
= \delta(t) \delta(y - y').
\label{green}
\ee
In what follows, 
our main task will be to find the brane black hole Green function.
Using the Fourier transform \cite{lea86}
$\tilde  
G(y, y';\omega) = \int_{0^{-}}^{\infty} dt~ G(y, y';t) e^{i \omega t}$ one can
reduce equation
(\ref{green}) to an ordinary differential equation.
The Fourier's transform is well defined for $Im~ \omega \ge 0$, while the 
corresponding inverse transform yields
\be
G(y, y';t) = {1 \over 2 \pi} \int_{- \infty + i \ep}^{\infty + i \ep}
d \omega~
\tilde G(y, y';\omega) e^{- i \omega t},
\ee
for some positive number $\ep$.
By virtue of the above
the Fourier's component of the Green's function $\tilde  G(y, y';\omega)$
can be written in terms of two linearly independent solutions for
homogeneous equation. Namely, one has
\be
\bigg(
{d^2 \over dy^2} + \omega^2 - V \bigg) \psi_{i} = 0, \qquad i = 1, 2.
\label{wav}
\ee
As far as 
the boundary conditions for $\psi_{i}$ are concerned they are described by purely ingoing waves
crossing the outer horizon $H_{+}$ of the 
brane black hole
$\psi_{1} \simeq e^{- i \omega y}$ as $y \rightarrow  - \infty$. On the other hand, 
$\psi_{2}$ should be damped exponentially at $i_{+}$, namely
$\psi_{2} \simeq e^{- \sqrt{m^2 - \omega^2}y}$ at $y \rightarrow \infty$.
%%%%%%%%%%%%%%%%%%%%%%%%%%%%%%%%%%%%%%%%%%%%%%%%%%%%%%%%%%%%%%%%%%%%%%%%%%%%%%%%%%%%%%%
A convenient form of the equation of motion for massive scalar field can be obtained by the transformation
as follows:
\be
\psi_{i} = {\bigg( 1 - {3M \over 2r} \bigg)^{1 \over 4}
\over \bigg( 1 - {2M \over r} \bigg)^{1/2}
\bigg( 1 - {\gamma M \over 2r} \bigg)^{1/4}} \xi,
\ee
where $i = 1,2$. 
On expanding Eq.(\ref{wav}) in a power  series of $ M/r$ neglecting terms of order
$\cO ((M/r)^2)$ and higher we finally achieve the relation
\be
{d^2 \over dr^2} \xi + \bigg[
\omega^2 - m^2 + {\omega^2 ({M \over 2}(5 + \gamma))
 - m^2 ({M \over 2}(1 + \gamma)) \over r}
 -{l(l + 1) \over r^2}
\bigg] \xi = 0.
\ee
The two basic solutions which are needed to construct the Green function, with the condition that
$\mid \omega \mid \ge m$ are given by $\tpsi_{1} = M_{\kappa, \tim}(2 \tom r)$ and $\tpsi_{2} = W_{\kappa, \tim}(2 \tom r)$,
with the following parameters:
\be
\tim = \sqrt{ 1/4 + l(l + 1)}, \qquad    \kappa = {\omega^2 ({M \over 2}(5 + \gamma) )
 - m^2 ({M \over 2}(1 + \gamma)) 
\over 2 \tom } \qquad        
\tom^2 = m^2 - \omega^2.
\label{casea}
\ee
Consequently it leads to the following spectral Green function:
\ben
G_{c}(y,y';t) &=& {1 \over 2 \pi} \int_{-m}^{m}dw
\bigg[ {\tpsi_{1}(y, \tom e^{\pi i})~\tpsi_{2}(y',\tom e^{\pi i}) \over W(\tom e^{\pi i})}
- {\tpsi_{1}(y, \tom )~\tpsi_{2}(y',\tom ) \over W(\tom )} 
\bigg] ~e^{-i w t} \\ \nonumber
&=& {1 \over 2 \pi} \int_{-m}^{m} dw f(\tom)~e^{-i w t}
\een 
First we shall take into account
the intermediate asymptotic behaviour of the massive scalar field. The range of parameters are 
$M \ll  r \ll t \ll M/(m M)^2$.
The intermediate asymptotic contribution to the Green function integral gives the frequency equal to 
$\tom = {\cO (\sqrt{m/t})}$, which in turns implies that $\kappa \ll 1$. Having in mind that $\kappa$ 
results from the $1/r$ term in the massive scalar field equation of motion, it depicts
the effect of backscattering off the spacetime curvature and in the case under consideration
the backscattering is negligible. Finally we are left with the result
\be
f(\tom) = {2^{2 \tim -1} \Gamma(-2\tim)~\Gamma({1 \over 2} + \tim) \over
\tim \Gamma(2 \tim)~\Gamma({1 \over 2} - \tim)} \bigg[
1 + e^{(2 \tim + 1) \pi i} \bigg]
(r' r)^{{1 \over 2} + \tim} \tom^{2 \tim},
\ee
where we have used the fact that $\tom r \ll 1$ and the form of $f(\tom)$
can be approximated by means of the fact that $M(a, b, z) = 1$ as $z$ tends to zero.
The resulting Green function reduces to the following:
\be
G_{c}(r, r';t) = {2^{3 \tim - {3\over2}} \over \tim \sqrt{\pi}}
{\Gamma(-2\tim)~\Gamma({1 \over 2} + \tim) \Gamma(\tim +1 ) \over
\tim \Gamma(2 \tim)~\Gamma({1 \over 2} - \tim)}
\bigg( 1 + e^{(2 \tim + 1) \pi i} \bigg)~(r r')^{{1 \over 2} + \tim} 
~\bigg( {m\over t} \bigg)^{{1 \over 2} + \tim}~J_{{1 \over 2} + \tim}(mt).
\ee  
Taking into account the limit when $t \gg 1/m$ one draws the conclusion that the spectral Green function implies
\be
G_{c}(r, r';t) = {2^{3 \tim - 1} \over \tim \sqrt{\pi}}
{\Gamma(-2\tim)~\Gamma({1 \over 2} + \tim) \Gamma(\tim +1 ) \over
\tim \Gamma(2 \tim)~\Gamma({1 \over 2} - \tim)}
\bigg( 1 + e^{(2 \tim + 1) \pi i} \bigg)~(r r')^{{1 \over 2} + \tim} 
~m^{\tim}~ t^{- 1 - \tim}~\cos(mt - {\pi \over 2}(\tim + 1)).
\label{gfim}
\ee  
Eq.(\ref{gfim}) depicts the oscillatory inverse power-law behaviour. In our case the intermediate
times of the power-law tail depends only on $\tim$ which in turn is a function of the multiple
moment $l$.
\par
However, the different pattern of decay is expected when  $\kappa \gg 1$, for the late-time
behaviour, when the backscattering off the curvature is important.
Consequently, $f(\tom)$ when $\kappa \gg 1$ may be rewritten in the following form:
\ben \label{fer}
f(\tom) &=& {\Gamma(1 + 2\tim)~\Gamma(1 - 2\tim) \over 2 \tim}~(r r')^{1 \over 2}
\bigg[ J_{2 \tim} (\sqrt{8 \kappa \tom r})~J_{- 2 \tim} (\sqrt{8 \kappa \tom r'})
- I_{2 \tim} (\sqrt{8 \kappa \tom r})~I_{- 2 \tim} (\sqrt{8 \kappa \tom r'}) \bigg] \\ \nonumber
&+&
{(\Gamma(1 + 2\tim))^2~\Gamma(- 2\tim)~\Gamma( {1 \over 2} + \tim - \kappa)
 \over 2 \tim ~\Gamma(2 \tim)~\Gamma({1 \over 2} - \tim - \kappa) }~(r r')^{1 \over 2}
~\kappa^{- 2 \tim}
\bigg[
J_{2 \tim} (\sqrt{8 \kappa \tom r})~J_{2 \tim} (\sqrt{8 \kappa \tom r'})
\\ \nonumber
&+& e^{(2 \tim + 1)}
I_{2 \tim} (\sqrt{8 \kappa \tom r})~I_{2 \tim} (\sqrt{8 \kappa \tom r'}) 
\bigg],
\een
where we have used the limit $M_{\kappa, \tim}(2 \tom r) \approx
\Gamma (1 + 2 \tim) (2 \tom r)^{1 \over 2}~\kappa^{- \tim}~J_{\tim}(\sqrt{8 \kappa \tom r})$ \cite{abr70}.
The first part of the above Eq.(\ref{fer}) the late time tail is proportional to $t^{-1}$
and it occurs that we shall concentrate on
the second term of the right-hand side of Eq.(\ref{fer}). For the case when 
$\kappa \gg 1$ it can be brought to the form
written as
\be
G_{c~(2)}(r, r';t) = {M \over 2 \pi} \int_{-m}^{m}~dw~e^{i (2 \pi \kappa - wt)}~e^{i \phi},
\ee
where we have defined
\be
e^{i \phi} = { 1 + (-1)^{2 \tim} e^{- 2 \pi i \kappa} \over
 1 + (-1)^{2 \tim} e^{2 \pi i \kappa}},
\ee
while $M$ provides the relation as follows:
\be
M = {(\Gamma(1 + 2\tim))^2~\Gamma(- 2\tim) \over 2 \tim ~\Gamma(2 \tim) }~(r r')^{1 \over 2}
\bigg[
J_{2 \tim} (\sqrt{8 \kappa \tom r})~J_{2 \tim} (\sqrt{8 \kappa \tom r'})
+ I_{2 \tim} (\sqrt{8 \kappa \tom r})~I_{2 \tim} (\sqrt{8 \kappa \tom r'}) 
\bigg].
\ee
At very late time both terms $e^{i w t}$ and $e^{2 \pi \kappa}$ are rapidly
oscillating. It means that the scalar waves are mixed states consisting of the states 
with multipole phases backscattered by spacetime curvature. Most of them cancel
with each others which have the inverse phase. In such a case, one can find the value of 
$G_{c~(2)}$ by means of the saddle point method. It could be found that the value $2 \pi \kappa
- wt$ is stationary at the value of $w$ equal to the following:
\be
a_{0} = \bigg[ { \pi~\bigg( \omega^2 ({M \over 2}(5 + \gamma) )
 - m^2 ({M \over 2}(1 + \gamma)) \bigg)
 \over 2 \sqrt{2} m} \bigg]^{1 \over 3}.
\ee
By virtue of saddle point method, on evaluating the adequate expressions, we derive the resultant spectral Green function
\be            
G_{c}(r, r';t) =  2\sqrt{6} m^{2/3} (\pi)^{5 \over 6}
\bigg[ 2Mm^2 
\bigg]^{1 \over 3}~
(mt)^{-{ 5 \over 6}}~\sin(mt)~\tpsi(r, m)~\tpsi(r', m),
\ee
One can observe that the dominant role in the late-time behaviour plays the term
proportional to $- 5/6$.
\par
The other case can be obtained if one 
considers $B(r) = \bigg( 1 - {2 \ga M \over r} \bigg)$
for the other model of brane black hole. It implies the following line element:
\be
ds^2 = - {1 \over \ga^2} \bigg( \ga -1 + \sqrt{1 - {2 \ga M \over r}} \bigg)^2 dt^2 +
{dr^2 \over \bigg( 1 - {2 \ga M \over r} \bigg)} + r^2 d \Omega^2.
\ee
For this case $\ga = 2 \beta - 1$. For $\beta = 1$ one has also the Schwarzschild limit.
\par
It is convenient to introduce the coordinate changes in the manner
\be
\psi_{i} = {\ga^{1 \over 2} \xi \over
\bigg( \ga - 1 + \sqrt{1 - {2 \ga M \over r}} \bigg)^{1 \over 2}
\bigg(1 - {2 \ga M \over r} \bigg)^{1 \over 4}},
\ee
where $i = 1,2$,
Let us expand Eq.(\ref{wav}) as a power  series of $ M/r$ neglecting terms of order
$\cO ((M/r)^2)$ and higher. It yields
\be
{d^2 \over dr^2} \xi + \bigg[
\omega^2~ \ga^2~\rho^2
- m^2 + {\omega^2 (4 \ga M (1 + \rho^2)) - m^2~2 M \ga
\over r}
 -{l(l + 1) \over r^2}
\bigg] \xi = 0,
\label{whita}
\ee
where $\rho^2 = (\ga - 1)^2 + 3$.\\
Eq.(\ref{whita}) can be brought to the form of the Whittaker's one with
the two basic solutions are needed to construct the Green function, with the condition that
$\mid \omega \mid \ge m$, i.e., $\tpsi_{1} = M_{\kappa, \tim}(2 \tom r)$ and $\tpsi_{2} = W_{\kappa, \tim}(2 \tom r)$.
The parameters of Whittaker's functions implies
\be
\tim = \sqrt{ 1/4 + l(l + 1)}, \qquad    \kappa = {\omega^2 (4 \ga M (1 + \rho^2)) - m^2~2 M \ga
\over 2 \tom}
 \qquad        
\tom^2 = m^2 - \omega^2~\ga^2~\rho^2.
\ee
It is easy to see from preceding case that 
the stationarity of $2 \pi \kappa - \omega t$ can be obtained for the parameter equal to 
\be
a_{0} = \bigg[ { \pi~\bigg( \omega^2 ({M \over 2}(5 + \gamma) )
 - m^2 ({M \over 2}(1 + \gamma)) \bigg)
 \over 2 \sqrt{2} m} \bigg]^{1 \over 3}.
\ee
Finally, the asymptotic late-time Green function reads
\be
G_{c}(r, r';t) =  {2\sqrt{6} m^{2/3} (\pi)^{5 \over 6} \over \ga^2 [(\ga - 1)^2 +3 ]}
\bigg[ m^2 (4 \ga M (\ga - 1)^2 + 14 \ga M) 
\bigg]^{1 \over 3}~
(mt)^{-{ 5 \over 6}}~\sin(mt)~\tpsi(r, m)~\tpsi(r', m).
\ee
One can observe that as in previous cases the dominant role in the asymptotic late-time decay of SI scalar 
hair in the spacetime of brane black hole plays the oscillatory tail with the decay rate proportional to
$t^{-5/6}$.

%%%%%%%%%%%%%%%%%%%%%%%%%%%%%%%%%%%%%%
\subsection{Dadhich-Maartens-Papadopoulous-Rezania (DMPR) brane black hole solution}
One can also mention the other case of the static spherically symmetric black hole localized on a three-brane
in five-dimensional gravity in Randall-Sundrum model \cite{ran99}.
Taking into account the effective field equations on the brane one gets the following brane black hole metric
\cite{dad00}:
\be
ds^2 = - \bigg( 1 - {2M \over M_{p}^2~r} + {q^2 \over \tM_{p}^2~r^2} \bigg) dt^2 +
{dr^2 \over \bigg( 1 - {2M \over M_{p}^2~r} + {q^2 \over \tM_{p}^2~r^2} \bigg)} + r^2~d\Omega^2,
\label{dila}
\ee
where $q$ is a dimensionless tidal parameter arising from the projection onto the brane of the gravitational
field in the bulk, $\tM_{p}$ is a fundamental five-dimensional Planck mass while $M_{p}$ is the effective Planck mass
in the brane world. Typically, one has that $\tM_{p} \ll M_{p}$. 
 In what follows we shall concentrate on the negative tidal charge which is claimed \cite{dad00}
to be the more natural case. Thus, the roots of $g_{00} = 0$ are respectively $r_{+}$ and $r_{-}$, namely
\be
r_{\pm} = {M \over \tM_{p}^2 }\bigg( 
1 \pm \sqrt{1 - {q M_{p}^4 \over M^2~\tM_{p}^2}} 
\bigg) =
M \bigg( 1 \pm \sqrt{1 + {Q \over M^2}} \bigg),
\ee
where $q = Q~\tM_{p}^2$.

\par
Let us assume further that the observer and the initial data are situated far away from the considered
black hole. On evaluating what follows, it is convenient to make the change of variables. Namely, it implies
\be
\psi_{i} = {\xi \over \bigg( 1 - {r_{+} \over r} \bigg)^{1/2}
\bigg( 1 - {r_{-} \over r} \bigg)^{1/2}},
\ee
where $i = 1,2$. In terms of the new variables
we expand Eq.(\ref{wav}) as a power  series of $ r_{\pm}/r$ neglecting terms of order
$\cO ((r_{\pm}/r)^2)$ and higher. Thus it provides the following:
\be
{d^2 \over dr^2} \xi + \bigg[
\omega^2 - m^2 + {(2 \omega^2 - m^2)( r_{+} + r_{-}) \over r} -{l(l + 1) \over r^2}
\bigg] \xi = 0.
\label{whitt}
\ee
Equation (\ref{whitt}) may be solved by means of Whittaker's functions. 
Just two basic solutions are needed to construct the Green function, with the condition that
$\mid \omega \mid \ge m$, i.e., $\tpsi_{1} = M_{\kappa, \tim}(2 \tom r)$ and $\tpsi_{2} = W_{\kappa, \tim}(2 \tom r)$.
The parameters of them are given in the manner
\be
\tim = \sqrt{ 1/4 + l(l + 1)}, \qquad    \kappa =  ( r_{+} + r_{-})
\bigg(
{m^2 \over 2 \tom} - \tom \bigg), \qquad        
\tom^2 = m^2 - \omega^2.
\label{aaa}
\ee
\par
The preceding section arguments can be repeated providing the form of $G_{c}(r, r';t)$
for the intermediate late-time decay of massive scalar hair with new Whittaker's function parameters
given by relation (\ref{aaa}). 
One can also see that the intermediate late-time decay of hair does not depend on brane black hole parameters but
it depends only on the scalar field mass and the multipole moment $l$.
As far as the late-time behaviour is concerned 
the spectral Green function for the late-time
behaviour of massive scalar field in the brane black hole spacetime can be obtained
using reasoning presented in 
Ref.\cite{rog07}, putting coupling constant in dilaton gravity $\alpha = 1$ and taking the exact values of $r_{+}, r_{-}$. 
Namely it can be written as
\be            
G_{c}(r, r';t) = { \sqrt{2} m^{4/3}} (\pi)^{5 \over 6}(M)^{1 \over 3}
(mt)^{-{ 5 \over 6}}~\sin(mt)~\tpsi(r, m)~\tpsi(r', m),
\ee

%%%%%%%%%%%%%%%%%%%%%%%%%%%%%%%%%%%%%%%%%%%%%%%%%%%%%%%%%%%%%%%%%%%%%%%%%%%%%%%%%%%%%%%%%%%%%%%%
\section{Conclusions}
In our paper we have elaborated the problem of the asymptotic tail behaviour of self-interacting
scalar fields in the background of various types of brane black holes. We are interested what kind
of mass-induced behaviours play the dominant role in the asymptotic intermediate and late-time
tails result from the decay of massive hair on the considered brane black holes.
In our research we took two kinds of brane black hole solutions presented in Refs.\cite{dad00} and \cite{cas02}.
In the case of intermediate asymptotic behaviour we obtained the oscillatory power law dependence which
in turn depend on the multiple number of the wave mode as well as the filed parameter $m$. However, as
in ordinary black hole analysis, this is not the final pattern of decay rate. The decay rate which is
the same for all kinds of black holes occurs at very late times. It stems from the resonance backscattering off the 
spacetime curvature. This decay rate is independent on the angular momentum parameter $l$ as well as the mass 
of hair on the brane black hole. 
\par
Our main result is that the asymptotic late-time behaviuor is of the form $t^{-5/6}$, exactly the same as
for static spherically symmetric black holes in Einstein or modified Einstein gravity (it happened that in dilaton gravity with
arbitrary coupling constant $\alpha$ the late-time behaviour is of the same form \cite{rog07}). 
Having in mind the late-time decay rate of massive scalar fields in the background of static spherically
symmetric black holes related to Einstein-Maxwell theory or the low-energy string theory, this kind of behaviour
should be more or less expected. It will be not amiss to investigate the decay rate of {\it black hair}
connected with other spins. The investigations in this direction is in progress and will be published elsewhere.

%%%%%%%%%%%%%%%%%%%%%%%%%%%%%%%%%%%%%%%%%%%%%%%%%%%%%%%%%%%%%%%%%%%%%%%%%%%%%%%%%%%%%%%%%
\begin{acknowledgments}
This work was partially financed by the budget funds in 2007 as the research project.
\end{acknowledgments}
%%%%%%%%%%%%%%%%%%%%%%%%%%%%%%%%%%%%%%%%%%%%%%%%%%%%%%%%%%%%%%%%%%%%%%%%%%%%%%%%%%%%%%%%%%

%%%%%%%%%%%%%%%%%%%%%%%%%%%%%%%%%%%%%%%%%%%%%%%%%%%%%%%%%%%%%%%%%%%%%%%%%%%%%%%%%%%%%%%%%%%%%%%%
%\begin{appendix}

%\section{Irred   } 
%\label{irtf}
%\end{appendix}

%\begin{acknowledgments}
%MR was supported ny grant
%\end{acknowledgments}
%%%%%%%%%%%%%%%%%%%%%%%%%%%%%%%%%%%%%%%%%%%%%%%%%%%%%%%%%%%%%%%%%%%%%%%%%%%%%%%%%%%%%%%%%%
%%%%%%%%%%%%%%%%%%%%%%%%%%%%%%%%%%%%%%%%%%%%%%%%%%%%%%%%%%%%%%%%%%%%%%%%%%%%%%%%%%%%%%%%%%%%%%%%%%%%%%%
%%%%%%%%%%%%%%%%%%%%%%%%%%%%%%%%%%%%%%%%%%%%%%%%%%%%%%%%%%%%%%%%%%%
%%%%%%%%%%%%%%%%%%%%%%%%%%%%%%%%%%%%%%%%%%%%%%%%%%%%%%%%%%%%%%%%%%%%%%%%%%%%%%%%%
%%%%%%%%%%%%%%%%%%%%%%%%%%%%%%%%%%%%%%%%%%%%%%%%%%%%%%%%%%%%%%%%%%%%%%%%%%%%%%%%%


\begin{thebibliography}{99}
%
\def\cmp#1#2#3{{ Commun. Math. Phys.} {\bf #1}, #2 (#3)}
\def\lmp#1#2#3{{ Lett. Math. Phys.} {\bf #1}, #2 (#3)}
\def\hpa#1#2#3{{ Hell. Phys. Acta} {\bf #1}, #2 (#3)}
\def\grg#1#2#3{{ Gen. Rel. Grav.} {\bf #1}, #2 (#3)}
\def\pr#1#2#3{{ Phys. Rev.} {\bf #1}, #2 (#3)}
\def\prl#1#2#3{{ Phys. Rev. Lett.} {\bf #1}, #2 (#3)}
\def\prd#1#2#3{{ Phys. Rev. D} {\bf #1}, #2 (#3)}
\def\pl#1#2#3{{ Phys. Lett} {\bf #1}, #2 (#3)}
\def\pla#1#2#3{{ Phys. Lett. A} {\bf #1}, #2 (#3)}
\def\plb#1#2#3{{ Phys. Lett. B} {\bf #1}, #2 (#3)}
\def\prep#1#2#3{{ Phys. Reports} {\bf #1}, #2 (#3)}
\def\phys#1#2#3{{ Physica} {\bf #1}, #2 (#3)}
\def\jcp#1#2#3{{ J. Comput. Phys.} {\bf #1}, #2 (#3)}
\def\jmp#1#2#3{{ J. Math. Phys.} {\bf #1}, #2 (#3)}
\def\jpm#1#2#3{{ J. Phys. A: Math. Gen.} {\bf #1}, #2 (#3)}
\def\cpr#1#2#3{{ Computer Phys. Rept.} {\bf #1}, #2 (#3)}
\def\cqg#1#2#3{{ Class. Quantum Grav.} {\bf #1}, #2 (#3)}
\def\cma#1#2#3{{ Computers Math. Applic.} {\bf #1}, #2 (#3)}
\def\mc#1#2#3{{ Math. Compt.} {\bf #1}, #2 (#3)}
\def\apj#1#2#3{{ Astrophys. J.} {\bf #1}, #2 (#3)}
\def\apjs#1#2#3{{ Astrophys. J. Suppl.} {\bf #1}, #2 (#3)}
\def\acta#1#2#3{{ Acta Astronomica} {\bf #1}, #2 (#3)}
%%%%%%%%%%%%%%%%%%%%%%%%%%%%%%%%%%%%%%%%%%%%%%%%%%%%%%%%%%%%%%%%%%%%%%%%%%
\def\apl#1#2#3{{Ann. Physik. (Leipzig)} {\bf #1}, #2 (#3)}
\def\anp#1#2#3{{Ann. Phys. } {\bf #1}, #2 (#3)}
\def\sa#1#2#3{{ Sov. Astro.} {\bf #1}, #2 (#3)}
\def\sia#1#2#3{{ SIAM J. Sci. Statist. Comput.} {\bf #1}, #2 (#3)}
\def\aa#1#2#3{{ Astron. Astrophys.} {\bf #1}, #2 (#3)}
\def\mnras#1#2#3{{ Mon. Not. R. astr. Soc.} {\bf #1}, #2 (#3)}
\def\npb#1#2#3{{ Nucl. Phys. B} {\bf #1}, #2 (#3)}
\def\prsla#1#2#3{{ Proc. R. Soc. London, Ser. A} {\bf #1}, #2 (#3)}
\def\jhep#1#2#3{{ JHEP} {\bf #1}, #2 (#3)}
\def\nuc#1#2#3{{Nuovo Cimento B } {\bf #1}, #2 (#3)}
\def\ijmp#1#2#3{{Int. J. Mod. Phys. D} {\bf #1}, #2 (#3)}
\def\atmp#1#2#3{{Adv. Theor. Math. Phys.} {\bf #1}, #2 (#3)}
\def\ptps#1#2#3{{Prog. Theor. Phys. Suppl.} {\bf #1}, #2 (#3)}
\def\ptp#1#2#3{{Prog. Theor. Phys. } {\bf #1}, #2 (#3)}
\def\lmp#1#2#3{{Lett. Math. Phys. } {\bf #1}, #2 (#3)}
%
\def\hepph#1#2{{ hep-ph }{\bf #1} (#2)}
\def\hepth#1#2{{ hep-th }{\bf #1} (#2)}
\def\grqc#1#2{{ gr-qc }{\bf #1} (#2)}
\def\ibid#1#2#3{{ {\it ibid.} }{\bf #1}, #2 (#3)}
%
%%%%%%%%%%%%%%%%%%%%%%%%%%%%%%%%%%%%%%%%%%%%%%%%%%%%%%%%%%%%%%%%%%%%%%
\bibitem{dad00}
N.Dadhich, R.Maartens, P.Papadopoulos, and V.Rezania, \plb{487}{1}{2000}.
\bibitem{cas02}
R.Casadio, A.Fabbri, and L.Mazzacurati, \prd{65}{084040}{2002},\\
C.Germani and R.Maartens, \ibid{64}{124010}{2001}.
\bibitem{kod02}
H.Kodama, \ptp{108}{253}{2002}.
\bibitem{tan03}
T.Tanaka, \ptps{148}{307}{2003}.
\bibitem{sea05}
S.Seahra, \prd{71}{084020}{2005}.
\bibitem{gal06}
C.Galfard, C.Germani, and A.Ishibashi, \prd{73}{064014}{2006}.

\bibitem{pri72}
R.H.Price, \prd{5}{2419}{1972}.
\bibitem{gun94}
C.Gundlach, R.H.Price and J.Pullin, \prd{49}{883}{1994}.
\bibitem{bic72}
J.Bicak, \grg{3}{331}{1972}.
\bibitem{pir1}
S.Hod and T.Piran, \prd{58}{024017}{1998}.
\bibitem{pir2}
S.Hod and T.Piran, \prd{58}{024018}{1998}.
\bibitem{pir3}
S.Hod and T.Piran, \prd{58}{024019}{1998}.
\bibitem{bur97}
L.M.Burko, {\it Abstracts of plenary talks and contributed papers},
15th International Conference on General Relativity and Gravitation,
Pune, 1997, p.143, unpublished.
\bibitem{hod98}
S.Hod and T.Piran, \prd{58}{044018}{1998}.
\bibitem{ja}
H.Koyama and A.Tomimatsu, \prd{63}{064032}{2001}.
\bibitem{ja1}
H.Koyama and A.Tomimatsu, \prd{64}{044014}{2001}.
\bibitem{mod01a}
R.Moderski and M.Rogatko, \prd{63}{084014}{2001}.
\bibitem{mod01b}
R.Moderski and M.Rogatko, \prd{64}{044024}{2001}.
\bibitem{jin04}
J.L.Jing, \prd{70}{065004}{2004}.
\bibitem{jin05}
J.L.Jing, \prd{72}{027501}{2005}.
\bibitem{bur04}
L.M.Burko and G.Khanna, \prd{70}{044018}{2004}.
\bibitem{xhe06}
X.He and J.L.Jing, \npb{755}{313}{2006}.
\bibitem{kon06}                
R.A.Konoplya, C.Molina, and A.Zhidenko, \prd{75}{084004}{2007}.
\bibitem{unn}
G.W.Gibbons, D.Ida and T.Shiromizu, \ptps{148}{284}{2003},\\
G.W.Gibbons, D.Ida and T.Shiromizu, \prl{89}{041101}{2002},\\
G.W.Gibbons, D.Ida and T.Shiromizu, \prd{66}{044010}{2002},\\
M.Rogatko, \cqg{19}{L151}{2002},\\
M.Rogatko, \prd{67}{084025}{2003},\\
M.Rogatko, \prd{70}{044023}{2004},\\
M.Rogatko, \prd{71}{024031}{2005},\\
M.Rogatko, \prd{73}{124027}{2006}.
\bibitem{car03}
V.Cardoso, S.Yoshida and O.J.C.Dias, \prd{68}{061503}{2003}.
\bibitem{mod05}
R.Moderski and M.Rogatko, \prd{72}{044027}{2005}.
\bibitem{shi00}
T.Shiromizu, K.Maeda, and M.Sasaki, \prd{62}{024012}{2000}.
\bibitem{wil93}
C.M.Will, {\it Theory and Experiment in Gravitational Physics}, (Cambridge University Press,
Cambridge, England, 1993).
\bibitem{ran99}
L.Randall and R.Sundrum, \prl{83}{3370}{1999}.
\bibitem{lea86}
E.W.Leaver, \prd{34}{384}{1986}.
\bibitem{abr70}
{\it Handbook of Mathematical Functions}, edited by M.Abramowitz and I.A.Stegun, (Dover, New York, 1970).
\bibitem{rog07}
M.Rogatko, \prd{75}{104006}{2007}.

%%%%%%%%%%%%%%%%%%%%%%%%%%%%%%%%%%%%%%%%%%%%%%%%%%%%%%%%%%%%%%%%%%%%%%%%%%%%%%%%%%%%%%%%%%%%%%%%%
\end{thebibliography}
\end{document}